\documentclass[twocolumn]{svjour3}       
\smartqed  

\usepackage{graphicx}
\usepackage{amsmath}
\usepackage{amsfonts}
\usepackage{makeidx}
\usepackage[squaren,thickspace,thickqspace]{SIunits}
\usepackage[english]{babel}
\usepackage{textcomp}
\usepackage[numbers]{natbib}
\makeindex 

\newcommand{\be}{\begin{eqnarray}}
\newcommand{\ee}{\end{eqnarray}}

\newcommand{\caf}{\ensuremath{^{40}{\rm Ca}^{+}\, }}

\newcommand{\Beplus}{\ensuremath{{^9}{\rm Be}^{+} \,}}

\newcommand{\D}{\mathrm{d}}
\newcommand{\I}{\mathrm{i}}
\newcommand{\Exp}[1]{\mathrm{e}^{#1}}
\newcommand{\der}[3][1]{\ifthenelse{#1=1}{\frac{\D#2}{\D#3}}
	{\frac{\D^#1#2}{\D#3^#1}}}%
\newcommand{\pder}[3][1]{\ifthenelse{#1=1}
	{\frac{\partial #2}{\partial #3}}{\frac{\partial ^#1#2}{\partial #3^#1}}}%
\newcommand{\rbra}[1]{\left(#1\right)}
	\renewcommand{\phi}{\varphi}
\let\FPi\Pi	\renewcommand{\Pi}{\mathit{\FPi}}

\begin{document}\sloppy

\title{All-solid-state continuous-wave laser systems for ionization, cooling and quantum state manipulation of beryllium ions}

\titlerunning{All-solid-state continuous-wave laser systems for beryllium ions}        

\author{Hsiang-Yu Lo*\thanks{*H.-Y. Lo and J. Alonso have contributed equally to this work} \and Joseba Alonso* \and Daniel Kienzler \and Benjamin C. Keitch \and Ludwig E. de Clercq \and Vlad Negnevitsky  \and Jonathan P. Home}

\authorrunning{H.-Y. Lo, J. Alonso \emph{et al.}} 

\institute{Institute for Quantum Electronics, ETH Z\"urich, Schafmattstrasse 16, 8093 Z\"urich, Switzerland\\\email{alonso@phys.ethz.ch}}

\date{Received: date / Accepted: date}

\maketitle

\begin{abstract}
We describe laser systems for photoionization, Doppler cooling and quantum state manipulation of beryllium ions. For photoionization of neutral beryllium, we have developed a continuous-wave \unit{235}{nm} source obtained by two stages of frequency doubling from a diode laser at \unit{940}{nm}. The system delivers up to \unit{400}{mW} at \unit{470}{nm} and \unit{28}{mW} at \unit{235}{nm}. For control of the beryllium ion, three laser wavelengths at \unit{313}{nm} are produced by sum-frequency generation and second-harmonic generation from four infrared fiber lasers. Up to \unit{7.2}{W} at \unit{626}{nm} and \unit{1.9}{W} at \unit{313}{nm} are obtained using two pump beams at 1051 and \unit{1551}{nm}. Intensity fluctuations below \unit{0.5}{\%} per hour (during 8 hours of operation) have been measured at a \unit{313}{nm} power of \unit{1}{W}. These systems are used to load beryllium ions into a segmented ion trap.
\end{abstract}

\section{Introduction}

The hyperfine levels of singly-charged earth-alkali ions have proven to be interesting as qubit states due to lack of spontaneous decay and the availability of transitions with first-order insensitivity to fluctuations in the ambient magnetic field \cite{05Langer,07Lucas}. The energy spacing between such states typically lies in the gigahertz regime. This allows for both optical addressing via Raman transitions and microwave addressing if the ion trap is small enough that near-field effects can offer strong magnetic field gradients \cite{11Ospelkaus,13Allcock}. Beryllium (\Beplus) is an example of such a system, and is currently being used by a number of experimental groups \cite{05Langer,05Blythe,08Rosenband,12Schwarz,13Tan,13Ball}.

\begin{figure}
\resizebox{0.5\textwidth}{!}{
  \includegraphics{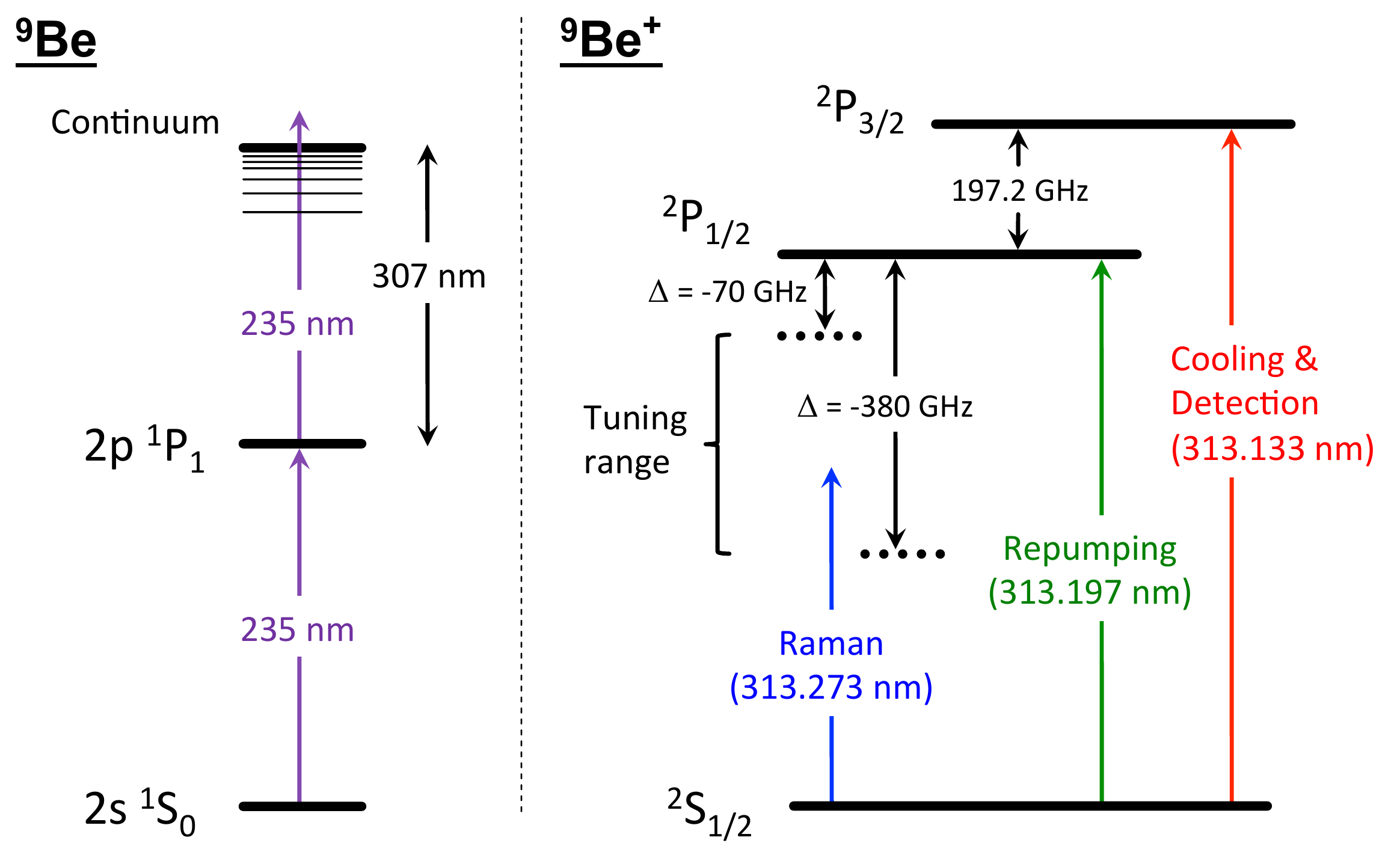}}
\caption{Relevant level schemes in beryllium. Left: two-photon ionization process by resonant excitation from $^1$S$_0$ to $^1$P$_1$ in neutral $^9$Be. Right: \Beplus energy levels addressed during trapping and manipulation.}
\label{fig:BeLvls}
\end{figure}

One of the major challenges of working with beryllium is that the lowest-energy transitions for excitation from the ground state of both the ion and the atom require wavelengths in the ultraviolet (UV) region of the spectrum, at 313 and \unit{235}{nm}, respectively (figure \ref{fig:BeLvls}). As a result, most  experiments performed until now have required frequency-doubled dye lasers for control of the ion. Photoionization of neutral beryllium has been previously performed using a frequency-doubled pulsed Titanium-Sapphire laser \cite{ThBlakestad}. These laser systems are expensive and complex, motivating the use of alternative approaches.

In this paper, we present a new setup for generating continuous-wave (CW) \unit{235}{nm} light for photoionization of beryllium atoms based on two stages of frequency doubling of light from a diode laser. In addition, for controlling beryllium ions we follow the approach demonstrated by Wilson \emph{et al.} \cite{11Wilson} to generate three desired UV wavelengths at \unit{313}{nm} starting from four infrared (IR) fiber-laser sources. The latter complements recent developments using telecom lasers at \unit{1565}{nm} \cite{11Vasilyev} or diode lasers at \unit{626}{nm} \cite{13Ball}, with the advantage that the powers we achieve are far higher.

\section{Laser source at 235~nm}

Ionization of neutral beryllium atoms requires an energy of \unit{9.3}{eV}, corresponding to a wavelength of \unit{133}{nm}. Since this wavelength lies in the vacuum UV, it is desirable to remove the electron using a two-photon excitation scheme. The photoionization cross-section is enhanced if the first stage of excitation is performed on resonance with an allowed transition in the neutral atom \cite{00Kjaergaard,03Lucas2}. Furthermore, resonance-enhanced photoionization is convenient in that it is species (and isotope) selective. For these reasons we choose to use two photons at \unit{235}{nm}, the first of which is resonant with the $^1$S$_0\leftrightarrow^1$P$_1$ transition of neutral beryllium (figure \ref{fig:BeLvls}, left). In what follows we describe our \unit{235}{nm} laser source, obtained by means of two stages of second-harmonic generation (SHG) starting from a commercial Toptica \unit{940}{nm} diode laser with a tapered amplifier (TA).

\begin{table*}\scriptsize
\caption{Details of the cavities for SHG of 470 and \unit{235}{nm} light. All mirror coatings were provided by Layertec GmbH. $w_\text{h,v}$ are the horizontal and vertical beam waists.}
\label{tab:235cavs}       
\begin{center}
\begin{tabular}{c c c c}
\hline
 & & 940$\rightarrow$470 nm & 470$\rightarrow$235 nm \\
\hline\hline
Crystal & Material & PPKTP & BBO \\
 & Phase-matching & QPM (type I) & CPM (type I) \\
 & Dimensions & $1\times2\times\unit{30}{mm^3}$ & $4\times4\times\unit{10}{mm^3}$ \\
 & Phase-matching angles $\theta/\phi$ & \unit{90/0}{\degree} & \unit{58.15/0}{\degree} \\
 & Surfaces & AR coated @ 940 nm & Brewster-cut (\unit{59.25}{\degree}) \\
 & Temperature stabilization & \unit{0.1}{\degree C} (@ \unit{21}{\degree C}) & Not stabilized \\
\hline
Mirrors & Radius of curvature M3 \& M4 & \unit{100}{mm} & \unit{38}{mm} \\
 & Reflectivity M1 & \unit{85.5}{\%} & \unit{99.0}{\%} \\
 & Coating M2-M4 & Sputter & Ion-beam sputter \\
\hline
Cavity & Long arm length & \unit{399}{mm} & \unit{320}{mm} \\
 & Crystal-mirror distance & \unit{50}{mm} & \unit{18}{mm} \\
 & Full opening angle ($\alpha$) & \unit{11.0}{\degree} & \unit{33.8}{\degree} \\
 & Mean waist short/long arm & \unit{50/212}{\micro m} & \unit{18/145}{\micro m} \\
 & Elipticity short/long arm ($w_\text{h}/w_\text{v}$) & 0.99/0.96 & 1.53/1.00 \\
 & Full spectral range ($FSR$) & \unit{560}{MHz} & \unit{805}{MHz} \\
 & Linewidth & \unit{43}{MHz} & \unit{2.7}{MHz} \\
 & Finesse ($\mathcal{F}$) & 13 & 300 \\
\hline
\end{tabular}
\end{center}
\end{table*}

\subsection{First SHG stage: 940 to 470~nm}

\begin{figure*}
\resizebox{0.75\textwidth}{!}{
  \includegraphics{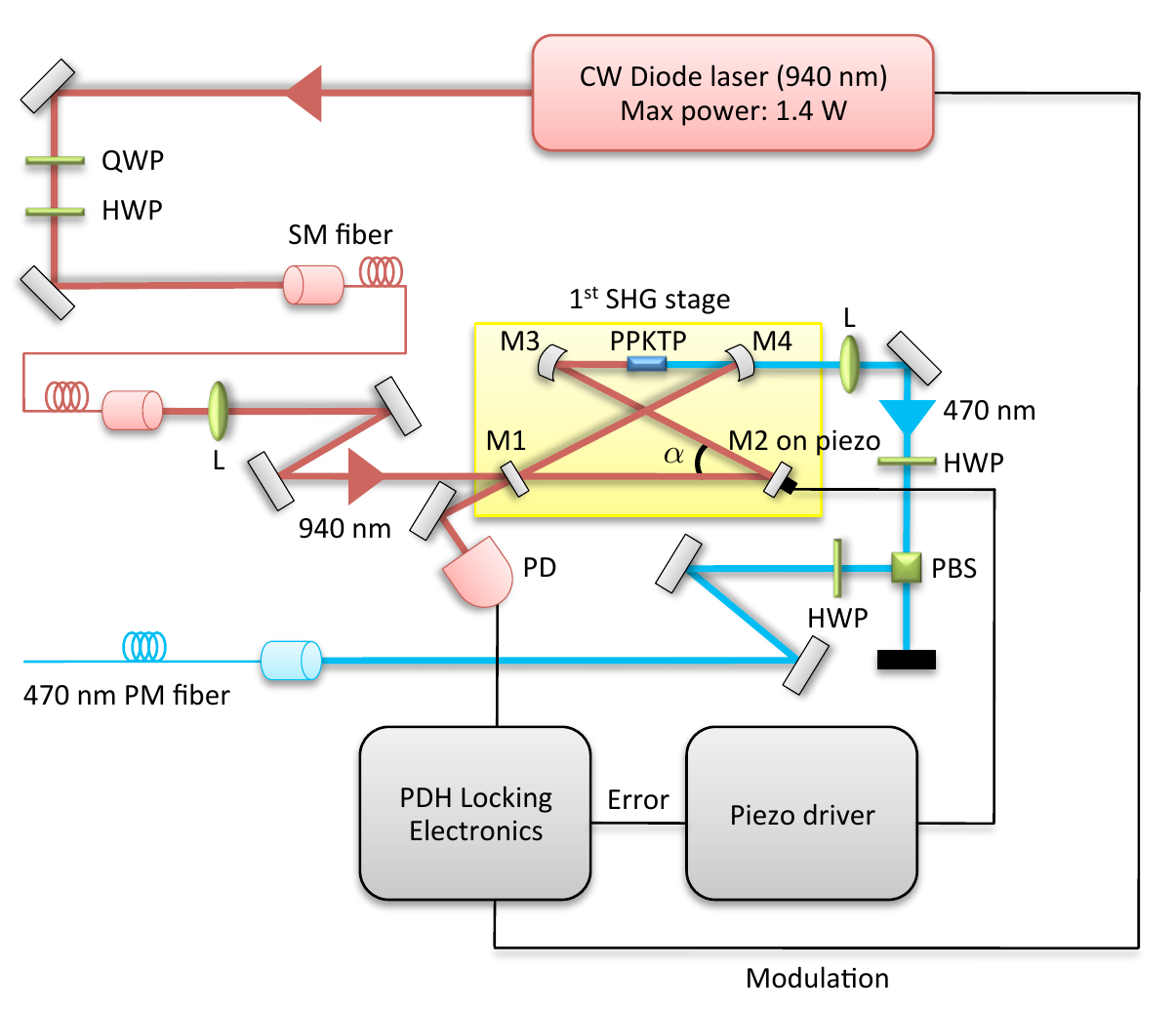}}
\caption{First SHG stage: 940 to 470~nm. Notation: L, lens; QWP, quarter-wave plate; HWP, half-wave plate; M$i$, cavity mirrors; PBS, polarizing beam-splitter; PD, photodiode; SM fiber, single-mode fiber; PM fiber, polarization-maintaining fiber; $\alpha$, full-opening angle. Also shown are the locking electronics. For details, see text.}
\label{fig:235setup1}
\end{figure*}

The first stage of frequency doubling converts 940 into \unit{470}{nm} light. A schematic of the experimental setup is shown in figure \ref{fig:235setup1}. The Toptica TA unit can deliver up to \unit{1.4}{W} at \unit{940}{nm}. We first pass this light through a single-mode (SM) fiber to clean up the spatial mode, resulting in up to \unit{700}{mW} which can be sent to a bowtie cavity for power build-up. The frequency conversion is performed using a periodically-poled potassium titanyl phosphate (PPKTP) crystal, chosen due to its high non-linearity and transparency at both 940 and \unit{470}{nm}. The PPKTP crystal (Raicol Crystal Ltd.) has a \unit{5.95}{\micro m} poling period and is temperature stabilized close to $\unit{21}{\degree C}$ for quasi-phase matching (QPM).

The design of the cavity's geometry is optimized to achieve high conversion efficiency while minimizing the dependence of the second-harmonic output power on the short arm optical length (M3-M4) \cite{11Wilson}. QPM is intrinsically free of walk-off (the Poynting and propagation vectors of the second-harmonic field are parallel), so a value for the Boyd-Kleinmann focusing parameter of $\xi=2.84$ should yield the maximum conversion efficiency \cite{68Boyd}. This parameter is defined as $\xi \equiv l_\text{c}/b$, with $l_\text{c}$ the crystal length and $b=2\pi n_\omega w^2/\lambda$ the confocal parameter of the laser beam. Here $n_\omega$ is the refractive index of the crystal sampled by the input light (of wavelength $\lambda$) and $w$ the waist size of the beam in the crystal. Based on previous observations of thermal lensing effects in a similar system operated at \unit{922}{nm} by Le-Targat \emph{et al.} \cite{05LeTargat}, we designed the cavity for a larger waist of $w=\unit{50}{\micro\meter}$ ($\xi=0.98$) to reduce the intensity of the generated blue light inside the crystal. The curved mirrors introduce astigmatism to the beam, which we minimize by using a small opening angle $\alpha\approx\unit{11}{\degree}$ (see figure \ref{fig:235setup1}), limited by geometric constraints imposed by the mirror-mounts. The transmission of the input coupling mirror M1 was chosen to be $T_1=\unit{14.5}{\%}$ in order to account for the expected cavity losses, and thus achieve an optimal impedance matching to the cavity for a predicted input power of $P_{940}=\unit{600}{mW}$. More details of the cavity design are given in table \ref{tab:235cavs}.

The length of the cavity is locked to the pump laser using the Pound-Drever-Hall (PDH) technique \cite{83Drever}. The \unit{940}{nm} diode is phase-modulated at \unit{15.6}{MHz} via a bias-tee input to the diode current. The light reflected from the cavity is collected by a fast photodiode and the signal is demodulated and filtered. The error signal thereby produced is used to feed back on the piezoelectric stack to which M2 is glued.

\begin{figure}
\resizebox{0.5\textwidth}{!}{
  \includegraphics{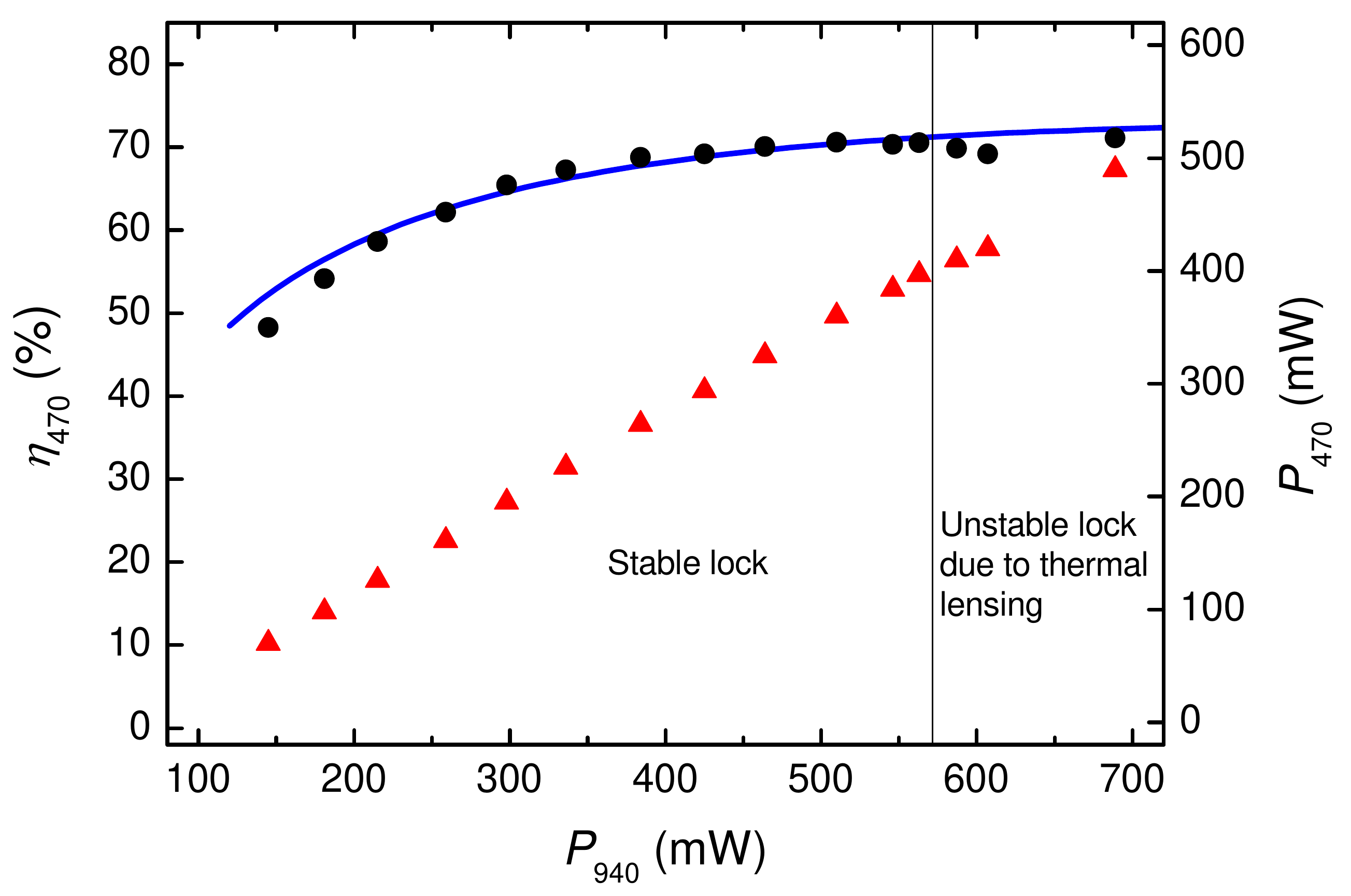}}
\caption{Measured net power at \unit{470}{nm} (triangles, referred to the right axis) and the power ratio of the 940$\rightarrow$\unit{470}{nm} doubling cavity (circles, referred to the left axis) as a function of the pump power. The solid line corresponds to the theoretical prediction of $\eta_{470}$ based on eq. \ref{eq:eta} in the appendix. To the right of the vertical line the cavity becomes unstable (see text).}
\label{fig:conveffvsPf}
\end{figure}

The results for this first stage of frequency doubling are shown in figure \ref{fig:conveffvsPf}, including both the second-harmonic output power $P_{470}$ and the ratio $\eta_{470}=P_{470}/P_{940}$ as a function of the input power $P_{940}$. The solid line results from the theoretical calculation using the theory given by Le-Targat \emph{et al.} \cite{05LeTargat}, which generalized the seminal work of Boyd and Kleinmann \cite{68Boyd} to a situation including an optical cavity \cite{01Freegarde}. Our calculations use the cavity parameters and the crystal properties provided by the manufacturer. These are displayed in tables \ref{tab:235cavs} and \ref{tab:gammatheo}. The theory is summarized in the appendix.

It is worth noting that, as was observed previously in a similar system operated at \unit{466}{nm} \cite{05LeTargat}, the output power is limited by thermal lensing induced by absorption of the \unit{470}{nm} light in the crystal. Our cavity becomes unstable for $P_{470}>\unit{400}{mW}$. For stable operation, we set $P_{940}\approx\unit{500}{mW}$ and obtain $P_{470}\approx\unit{350}{mW}$ ($\eta_{470}\approx\unit{70}{\%}$).

\subsection{Second SHG stage: 470 to 235~nm}

\begin{figure}
\resizebox{0.5\textwidth}{!}{
  \includegraphics{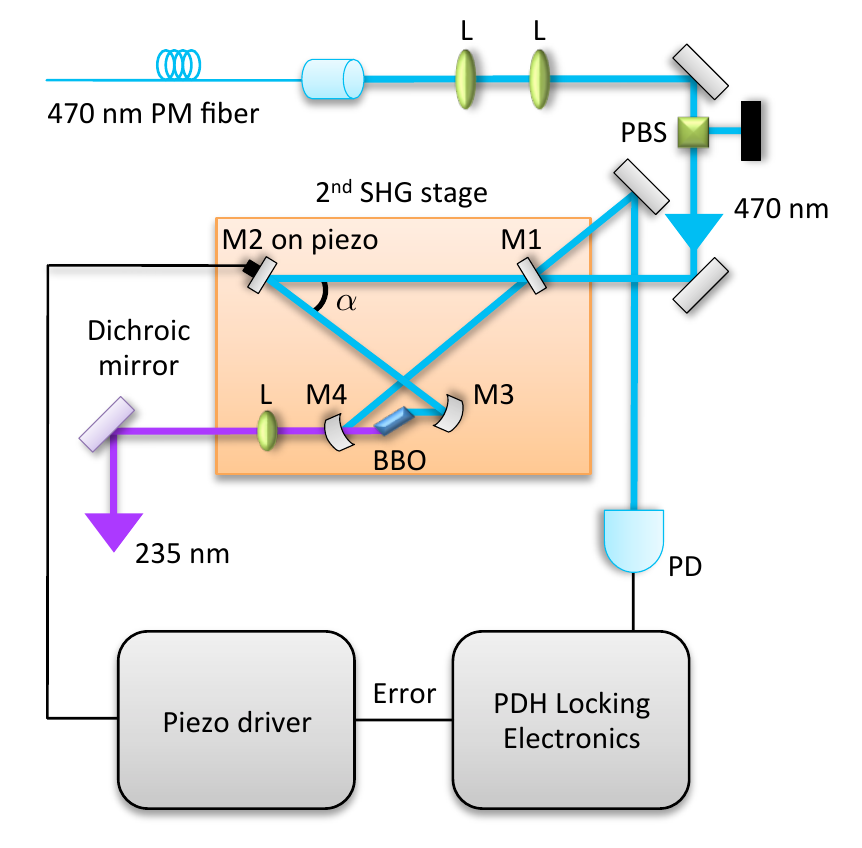}}
\caption{Second SHG stage: 470 to \unit{235}{nm}. Labels are as in figure \ref{fig:235setup1}. For details, see text.}
\label{fig:235setup2}
\end{figure}

The \unit{470}{nm} light generated in the first SHG stage is sent through a \unit{30}{m} long polarization-maintaining (PM) fiber (OZ Optics) to a neighboring laboratory. The output mode of the fiber is matched to the spatial mode of the second bowtie cavity with a two-lens telescope, and the polarization is cleaned using a polarizing beam-splitter (PBS) cube (see figure \ref{fig:235setup2}). Input-coupling losses and absorption in the fiber result in a maximum value of $\approx\unit{140}{mW}$ of light incident on the 470$\rightarrow$\unit{235}{nm} cavity.

For this SHG stage we use a BBO crystal (Castech Inc.) Brewster-cut for \unit{470}{nm} which achieves critical type-I phase matching (CPM) at an angle of \unit{58.15}{\degree}. This cavity was also designed to optimize the conversion efficiency of the frequency doubling while being insensitive to changes in the positions of the focusing mirrors M3 \& M4 \cite{11Wilson}. By contrast with the first frequency-doubling stage, the use of angle phase-matching intrinsically results in walk-off which limits the length of the conversion region of the crystal. The optimized cavity parameters are given in table \ref{tab:235cavs}. The length of the cavity is also stabilized using the PDH method in order to keep it on resonance with the pump laser. Since the  modulation frequency applied to the diode laser (\unit{15.6}{MHz}) is comparable to the linewidth of the 940$\rightarrow$\unit{470}{nm} cavity, the \unit{470}{nm} light also exhibits modulation sidebands at \unit{15.6}{MHz} and can be directly used for the PDH locking scheme.

\begin{figure}
\resizebox{0.5\textwidth}{!}{
  \includegraphics{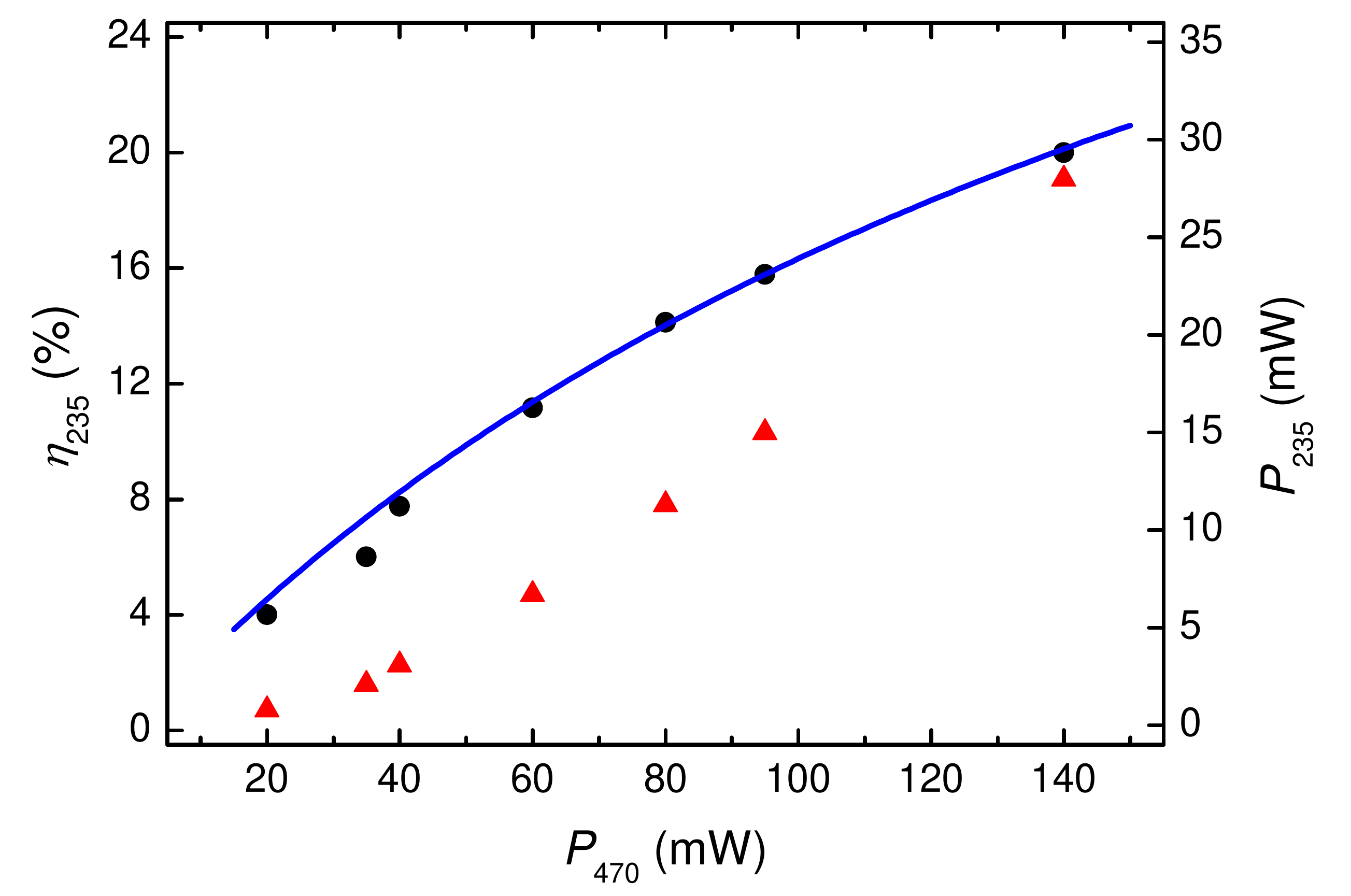}}
\caption{Measured net power at \unit{235}{nm} (triangles, referred to the right axis) and the power ratio of the 470$\rightarrow$\unit{235}{nm} doubling cavity (circles, referred to the left axis) versus the pump power at \unit{470}{nm}. The solid line corresponds to the theoretical prediction of $\eta_{235}$ based on eq. \ref{eq:eta} in the appendix.}
\label{fig:conveffvsPs}
\end{figure}

Figure \ref{fig:conveffvsPs} shows the UV output power $P_{235}$ along with the power ratio $\eta_\text{235}=P_{235}/P_{470}$ as a function of the input power $P_{470}$. Note that we define $P_{235}$ and $\eta_{235}$ in terms of the net second-harmonic power at the output of the cavity. With $P_{470}\approx \unit{140}{mW}$, we obtain $P_{235} \approx \unit{28}{mW}$ ($\eta_{235}\approx\unit{20}{\%}$). Considering the \unit{22}{\%} reflection at the output surface of the Brewster-cut BBO crystal, the total UV light generated is $\approx \unit{36}{mW}$. The theoretical curve is calculated using the values given in table \ref{tab:gammatheo}, yielding $\Gamma_\text{eff}\approx\unit{1.6\times10^{-4}}{W^{-1}}$. The level of agreement between the measured values of $\eta_\text{235}$ and those predicted by the theory is reasonable given the assumptions made in the theory, which include perfect phase-matching of a circular beam and no absorption in the crystal (see table \ref{tab:gammatheo}).

The counter-propagating fundamental mode of a bowtie cavity can be excited by any back-reflecting element in the cavity, which can lead to considerable back-circulating power if the finesse $\mathcal{F}$ of the cavity is high \cite{90Hemmerich}. In the 470$\rightarrow$\unit{235}{nm} cavity, we observe that light is emitted along a direction which is consistent with this effect. This leads to a reduction in $\eta_\text{235}$ and an unstable output power. The back-circulation could likely be eliminated by reducing the reflectivity of M1 if required \cite{90Hemmerich}. Since photoionization does not require high stability or high power, we have currently chosen not to pursue this.

The power we achieve is comparable to the average power of the pulsed laser used in \cite{ThBlakestad}, and considerably higher than values which have been previously used to perform photoionization of other atomic species \cite{00Kjaergaard,03Lucas2}. The $^1$S$_0\leftrightarrow^1$P$_1$ transition of neutral beryllium atoms has a  saturation intensity of \unit{8.9}{mW/mm^2}, corresponding to \unit{70}{\micro W} for a beam waist of \unit{50}{\micro m}.

\section{Laser sources at 313~nm}
\label{sec:313}

\begin{figure*}
\resizebox{1.0\textwidth}{!}{
\includegraphics{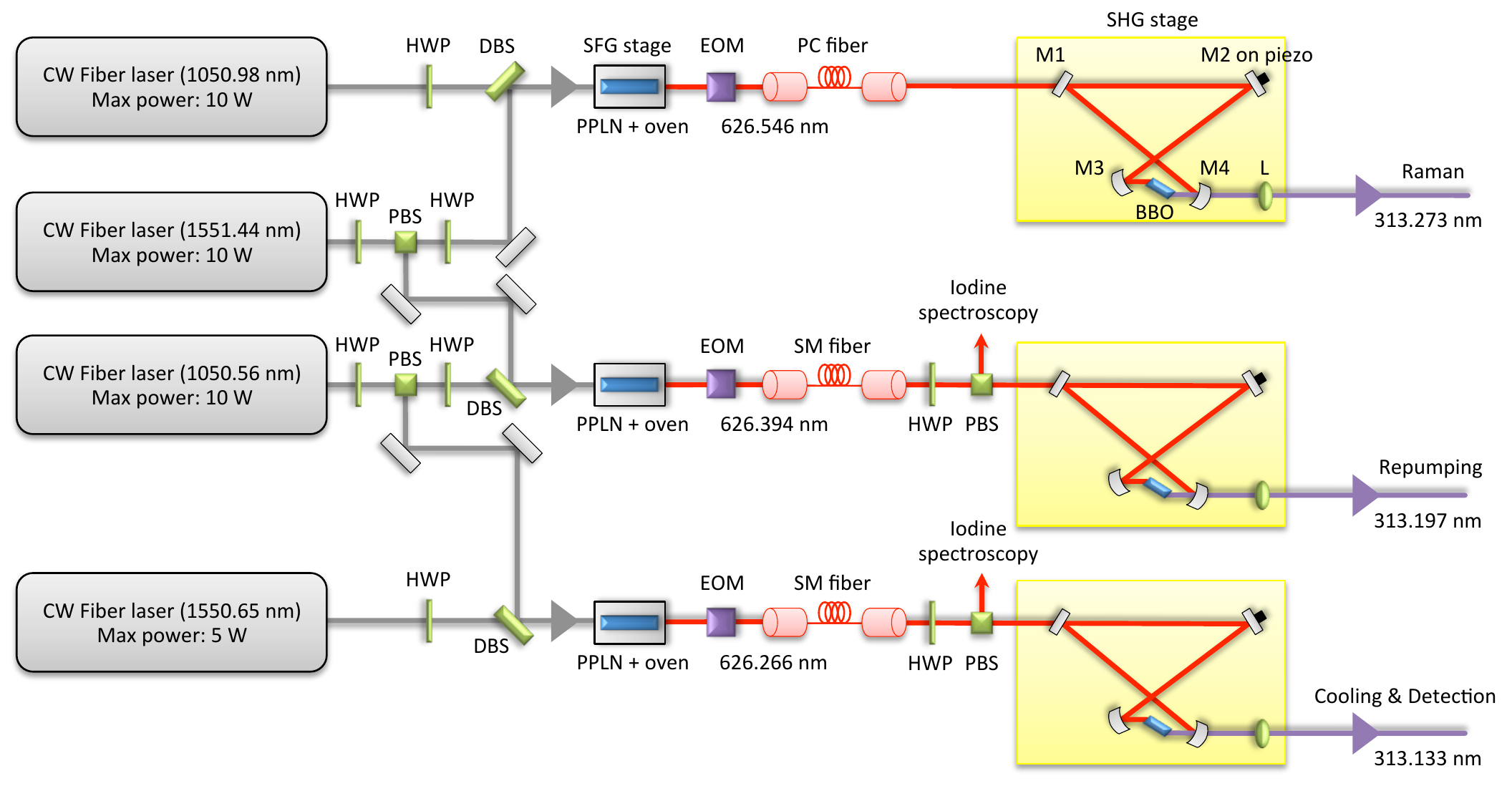}}
\caption{Setup for generating three wavelengths at \unit{313}{nm} for Doppler cooling, optical pumping and coherent manipulations of beryllium ions. Notation: DBS, dichroic beam-splitter; EOM, electro-optic modulator; PC fiber, photonic-crystal fiber; the rest are as in figure \ref{fig:235setup1}. The locking electronics and focusing lenses are not shown. The wavelengths indicated for the Raman setup are approximately in the middle of the range of operation. For details, see text.}
\label{fig:313setup}
\end{figure*}

Quantum control experiments using beryllium ions require near resonant light for Doppler cooling and optical pumping and, in our case, a far-detuned Raman laser for coherent manipulations \cite{98Wineland2}. As shown in figure \ref{fig:BeLvls} (right), the wavelengths of the $^2$S$_{1/2}\leftrightarrow^2$P$_{3/2}$ and $^2$S$_{1/2}\leftrightarrow^2$P$_{1/2}$ transitions are at \unit{313.133}{nm} and \unit{313.197}{nm}, respectively. These transitions have saturation intensities of \unit{0.8}{mW/mm^2} \cite{ThLanger}, corresponding to \unit{6}{\micro W} for a beam waist of \unit{50}{\micro m}. We have chosen to operate our Raman laser system at a wavelength between 313.221 and \unit{313.322}{nm}, detuned to the red of the $^2$P$_{1/2}$ manifold by between \unit{-380}{} and \unit{-70}{GHz}. Small detunings allow for lower power for a given Raman transition rate, but lead to more spontaneous scattering of photons which will induce errors in the quantum state manipulation. Therefore, for high-fidelity control of a quantum system, a large detuning from the excited state is preferable \cite{07Ozeri}. The laser systems to generate all three wavelengths are shown schematically in figure \ref{fig:313setup}.

\subsection{SFG stage: infrared to 626 nm}

To produce the three UV wavelengths we require, we start from four fiber lasers (NP Photonics) at \unit{1550.65}{nm}, \unit{1050.56}{nm}, \unit{1551.44}{nm} and \unit{1050.98}{nm} which can be thermally tuned over a range of \unit{80}{GHz} and fine-tuned over $\pm\unit{150}{MHz}$ by means of an internal piezoelectric element. The laser outputs (PM fibers) are terminated with an integrated beam-collimation system, producing beams of diameters between 3 and \unit{4}{mm}. The sum-frequency generation (SFG) is performed using magnesium-oxide-doped periodically-poled lithium niobate (MgO:PPLN) crystals (Covesion Ltd.). Each crystal contains three different poling periods of around \unit{11.5}{\micro m} on strips \unit{1}{mm} wide which run down the length of the \unit{4}{cm} long crystal. The PPLN crystals are maintained at a constant operating temperature in order to achieve the phase-matching condition for maximizing the conversion efficiency. In our case, we obtain maximal \unit{626}{nm} output power for temperatures close to $\unit{180}{\degree C}$, stabilized to within $\pm\unit{0.01}{\degree C}$. In order to access the highest non-linear coefficient of the crystal, the polarization axis of the light is parallel to the thickness of the crystal.

The \unit{1050.56}{nm} light is shared between the cooling and repumping setups (figure \ref{fig:313setup}). The power ratio can be tuned by rotating the half-wave plate in front of the PBS. The same mechanism applies for the \unit{1551.44}{nm} laser, which is split between the repumping and Raman setups. In all three setups, two of the fiber-laser outputs (pump beams) are overlapped using a dichroic element after passing through independent beam-shaping telescopes (not shown in the figure) that focus the beam to an optimal waist inside the crystal. The beam waist that maximizes the conversion efficiency was experimentally found to be \unit{58 \pm 5}{\micro m} at the center of the crystal. This is larger than the \unit{40}{\micro m} predicted from the Boyd-Kleinmann theory. A possible explanation for this difference could be phase mismatch or imperfect spatial-mode overlap arising from absorption-induced heating in the crystal \cite{98Batchko}.

\begin{figure}
\resizebox{0.5\textwidth}{!}{
  \includegraphics{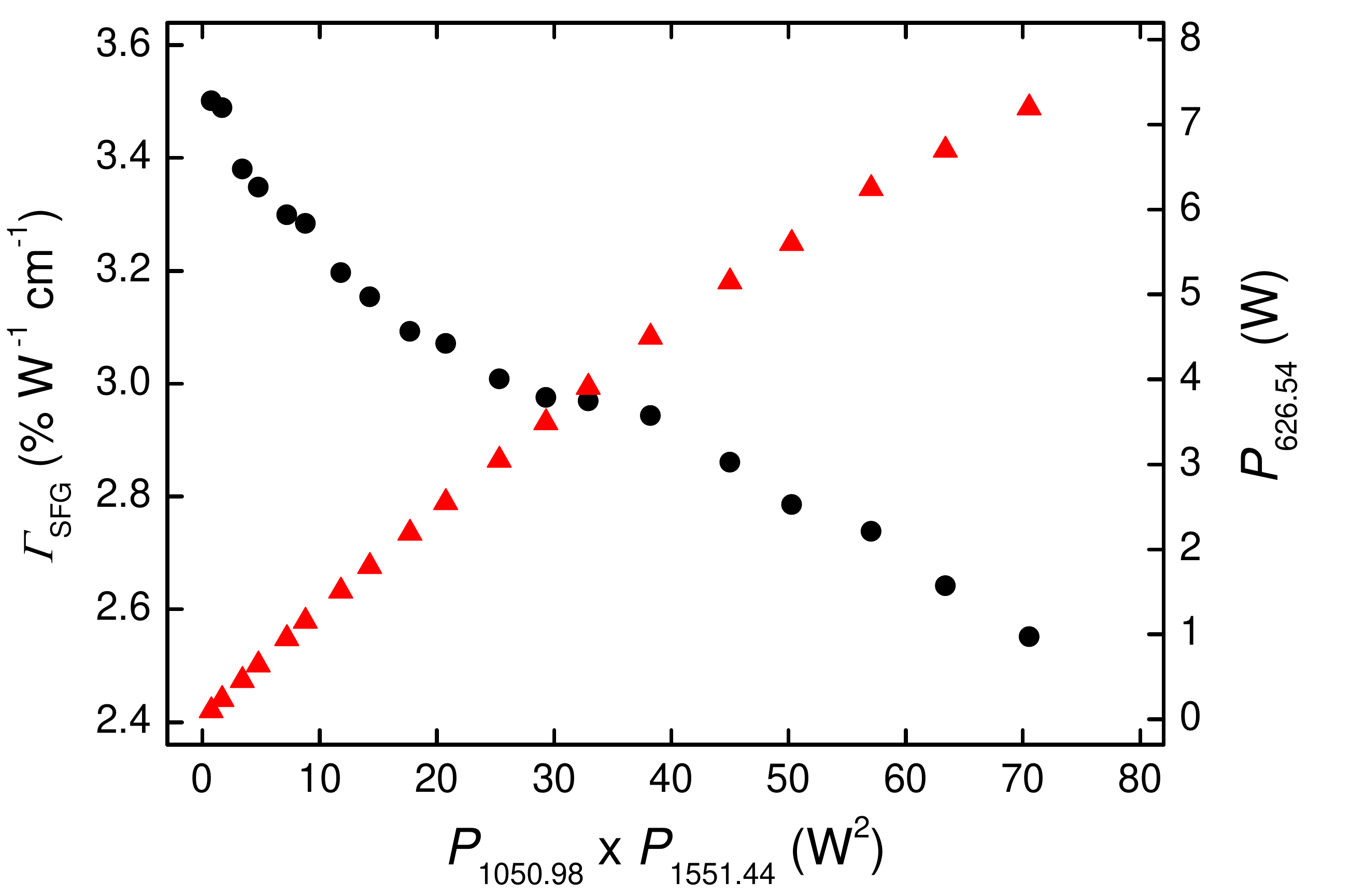}}
\caption{Net power of \unit{626}{nm} light measured at the output as a function of the product of pump powers (triangles, referred to the right axis). Also shown (circles, referred to the left axis) is the sum-frequency-generation efficiency per unit length of the crystal.}
\label{fig:SFG_Raman_Power}
\end{figure}

\begin{figure}
\resizebox{0.5\textwidth}{!}{
  \includegraphics{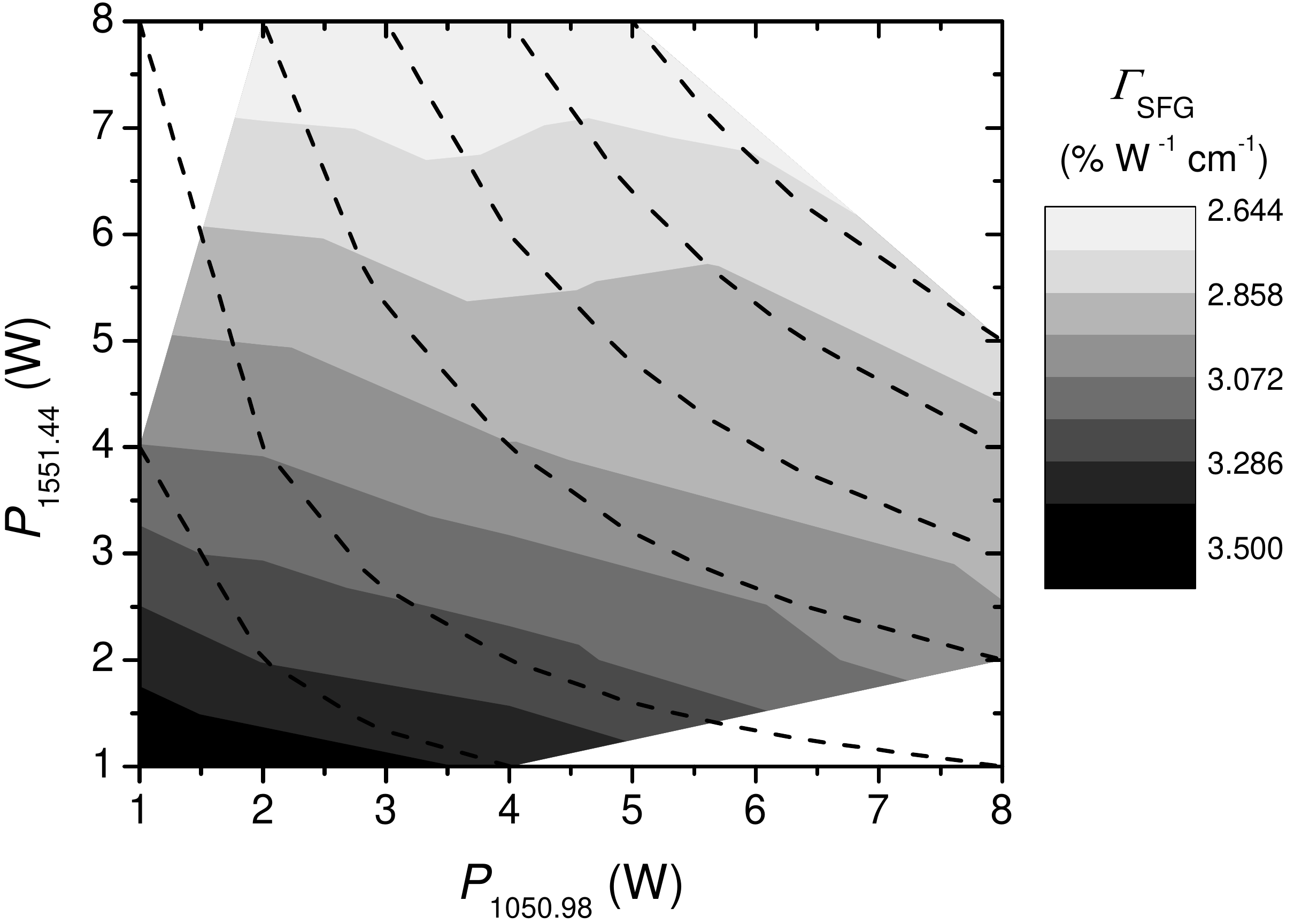}}
\caption{Conversion efficiency per unit length of the crystal measured for different pump powers. The dashed lines correspond to combinations of $P_{1551.44}$ and $P_{1050.98}$ with a constant product. The data points were taken along these curves.}
\label{fig:626powerdependent}
\end{figure}

All SFG setups exhibit similar performance. For the Raman light (SFG of the \unit{1050.98}{nm} and \unit{1551.44}{nm}) the output power for different products of the pump powers is plotted in figure \ref{fig:SFG_Raman_Power}. We have generated up to \unit{7.2}{W} of red light at \unit{626.54}{nm} using pump powers of \unit{8.5}{W} at \unit{1050.98}{nm} and \unit{8.3}{W} at \unit{1551.44}{nm}. Also shown is the conversion efficiency per unit length
\begin{equation}
\Gamma_\text{SFG} = \frac{P_{626}}{P_{1051}P_{1551}l_\text{c}},
\end{equation}
which characterizes the performance of the non-linear crystal. We find a maximum value of $ \Gamma_\text{SFG} \approx \unit{3.5}{\%~W^{-1}cm^{-1}}$, which is reduced to $\approx\unit{2.5}{\%~W^{-1}cm^{-1}}$ at the highest output power, possibly due to absorption-induced heating in the crystal. To test whether these effects are more prominent in the IR or the visible light, we measured $\Gamma_\text{SFG}$ for different combinations of pump powers with a constant product $P_{1050.98}\times P_{1551.44}$, which would be expected to produce the same amount of red light. The results (see figure \ref{fig:626powerdependent}) show that the conversion efficiency is lowest when $P_{1551.44} > P_{1050.98}$. The Pearson correlation coefficients $\mathcal{R}_{\lambda}$ between $\Gamma_\text{SFG}$ and the pump powers are $\mathcal{R}_{1050.98} \approx -0.5$ and $\mathcal{R}_{1551.44} \approx -0.9$, indicating that the absorption of the \unit{1551.44}{nm} light contributes most to thermal effects.

At an output power of $\approx\unit{7}{W}$ the drift was observed to be below \unit{0.2}{\%} per hour over 10 hours of operation. After running this system for more than a year at a variety of input and output powers, we have seen little change in the conversion efficiency, and no realignment of the optical components has been required.

Since the detection and repumping beams must provide light resonant with the $^2$S$_{1/2}\leftrightarrow^2$P$_{3/2}$ and $^2$S$_{1/2}\leftrightarrow^2$P$_{1/2}$ transitions, the absolute frequencies of these lasers must be actively stabilized. The transition linewidths are $\gamma/{2\pi} \approx \unit{19.4}{MHz}$, so we aim for frequency stability on the level of a megahertz. To that end, a small portion of the \unit{626}{nm} light is picked off and sent through an optical fiber to a Doppler-free saturated-absorption-spectroscopy setup with an iodine cell \cite{BkDemtroder,ThKing}. In this way the light can be locked to one of the molecular hyperfine features of the iodine vapor. Feedback control of the laser frequency is performed using the piezoelectric element in one of the fiber lasers. In the Raman setup, the stability of the absolute frequency is not as critical due to the large detuning from resonance, and the natural stability of the fiber lasers (which we have measured to drift by less than \unit{60}{MHz} over a day) is sufficient for our purposes.

\subsection{SHG stage: 626 to 313 nm}

The beams from the SFG setups are coupled into high-power SM fibers and sent to frequency doubling cavities for conversion to \unit{313}{nm}. The coupling efficiency is $\approx\unit{90}{\%}$ in the repumping and cooling setups (\unit{5}{m} long SM fibers from OZ Optics), owing to the high quality of the Gaussian modes produced by the fiber lasers, which is transferred to the \unit{626}{nm} light. The power transmitted through the fibers saturates at $\approx\unit{2}{W}$ due to stimulated Brillouin scattering \cite{BkAgrawal}. In order to avoid this problem in the Raman setup, where we require higher powers, we use a photonic-crystal fiber (NKT Photonics). The limitation is in this case the coupling efficiency into the fiber, which is $\approx\unit{60}{\%}$.

SHG from 626 to \unit{313}{nm} is performed in all three cases using a BBO crystal Brewster-cut for \unit{626}{nm} with a phase-matching angle of \unit{38.35}{\degree}. The crystals are placed in bowtie cavities ($FSR\approx\unit{845}{MHz}$ and $\mathcal{F}\approx275$) for pump-power enhancement. The beam waists in the BBO crystals are 36.7 \unit{(23.6)}{\micro m} for the horizontal (vertical) direction. The design and properties of the doubling cavities are very similar to those reported by \cite{11Wilson} except that we use a Pound-Drever-Hall rather than H{\"a}nsch-Coullioud scheme to stabilize the cavity length. In order to generate the required modulation for the PDH stabilization, we pass the \unit{626}{nm} beams through electro-optic phase modulators (Qubig) placed directly after the PPLN crystals and driven at frequencies closed to \unit{125}{MHz}. All of the SHG conversion setups are similar, so below we give explicit details for the Raman setup, for which we work with the highest powers.

\begin{figure}
\resizebox{0.5\textwidth}{!}{
  \includegraphics{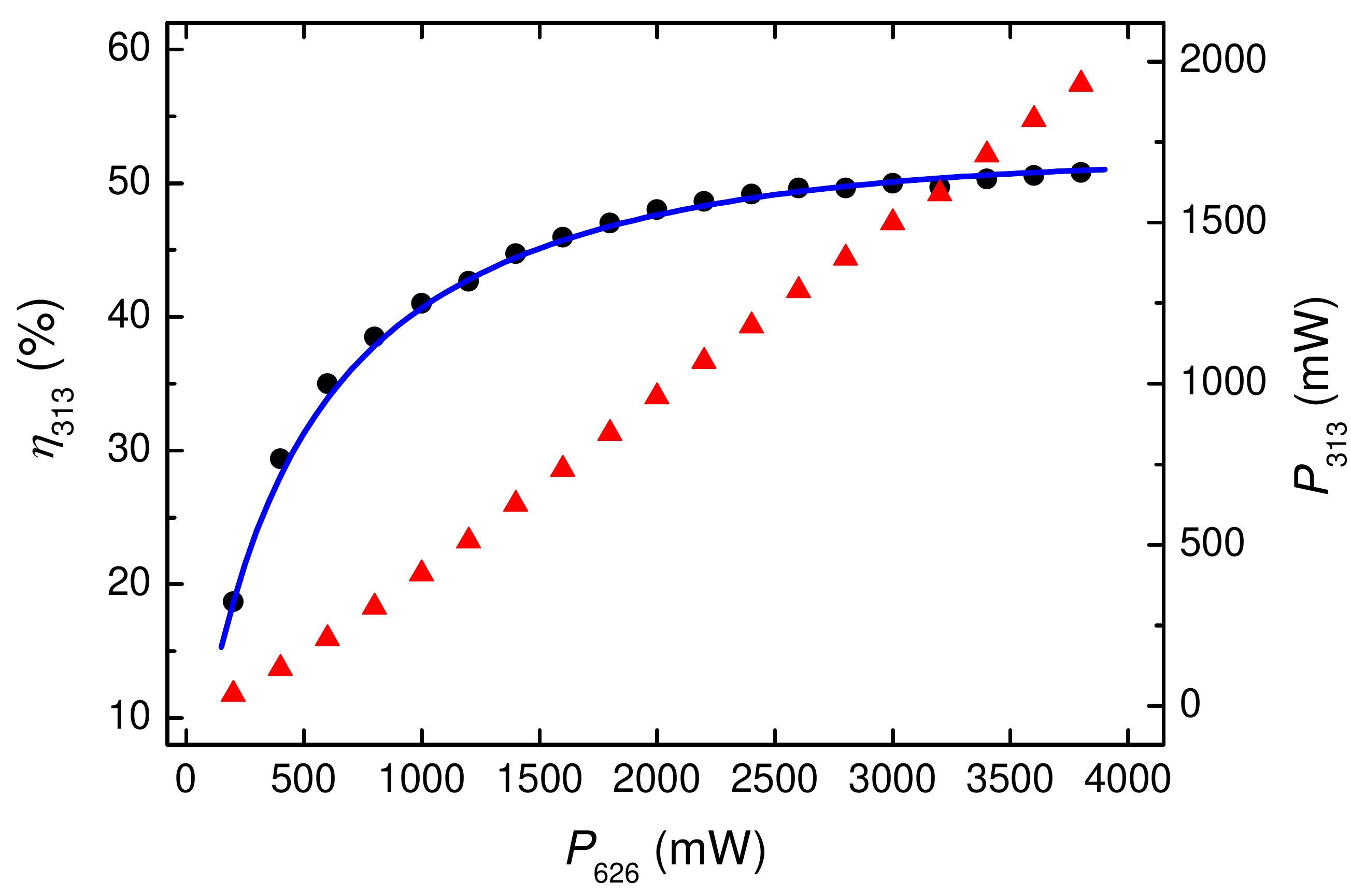}}
\caption{Measured net power at \unit{313}{nm} (triangles, referred to the right axis) and the power ratio of the 626$\rightarrow$\unit{313}{nm} Raman doubling cavity (circles, referred to the left axis) versus the input power at \unit{626}{nm}. The solid line corresponds to the theoretical prediction of $\eta_{313}$ based on eq. \ref{eq:eta} in the appendix.}
\label{fig:SHG_Raman_Power}
\end{figure}

Figure \ref{fig:SHG_Raman_Power} shows the \unit{313}{nm} output power $P_{313}$ and the power ratio $\eta_{313}=P_{313}/P_{626}$ as a function of the power $P_{626}$ coupled into the Raman doubling cavity. As before, $P_{313}$ and $\eta_{313}$ are defined in terms of the net second-harmonic power delivered by the system. The maximum output power obtained is $\approx \unit{1.95}{W}$, with $P_{626}\approx \unit{3.8}{W}$ ($\eta_{313}\approx\unit{52}{\%}$). Including the \unit{20}{\%} reflection at the output facet of the Brewster-cut BBO crystal, overall $\approx \unit{2.4}{W}$ at \unit{313}{nm} are produced. The parameters for the theoretical calculation of equation \ref{eq:eta} are given in table \ref{tab:gammatheo}.

In order to characterize the long-term performance of the UV light production, we have monitored the stability of the \unit{313}{nm} power. For these tests, output powers up to $P_{313}\approx\unit{1}{W}$ ($P_{626}\approx\unit{2.2}{W}$) were used. Observed drifts were below \unit{0.5}{\%} per hour during 8 hours of continuous operation.

\section{Ion loading}

\begin{figure}
\resizebox{0.4\textwidth}{!}{
  \includegraphics{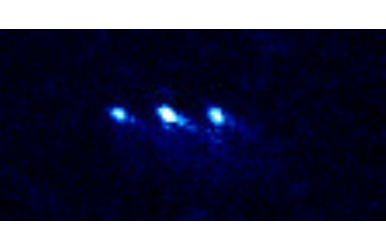}}
\caption{Trapped ion crystal with three \Beplus ions fluorescing at \unit{313}{nm}.}
\label{fig:Beions}
\end{figure}

To demonstrate the combined use of our photoionization and cooling laser systems, we load beryllium ions into a micro-fabricated segmented trap designed for both \caf and \Beplus ions (details of the trap will be given elsewhere). As a source for \Beplus production we make use of a beryllium wire tightly coiled around a tungsten wire. The tungsten is heated by running an electrical current through it, which in turn heats the beryllium up \cite{ThBlakestad}. Neutral atoms are desorbed in this way, and then collimated to travel through the trapping area envisioned for loading. There the neutral atom beam is shone simultaneously with \unit{235}{nm} light (for resonant ionization) and two right-circularly polarized \unit{313}{nm} beams. The first one is slightly red-detuned by $\approx\unit{10}{MHz}$ from the $^2$S$_{1/2}\leftrightarrow^2$P$_{3/2}$ ion transition for Doppler cooling. The second is driven at higher power and is red-detuned by $\approx\unit{600}{MHz}$ for efficient cooling during the loading process (when ions are most energetic) as well as repumping. The trapped ions are monitored throughout the process by means of a CCD-camera with which we image \Beplus fluorescence at \unit{313}{nm} (figure \ref{fig:Beions}). We typically load an ion after less than a minute.

\section{Conclusions}

We have constructed and characterized all-solid-state continuous-wave laser systems for photoionization loading, cooling and quantum state manipulation of beryllium ions. The wavelength required for photoionization of neutral beryllium atoms (\unit{235}{nm}) is created by two stages of second-harmonic generation (SHG) using PPKTP and BBO non-linear crystals placed inside resonant cavities. In the first stage, we have demonstrated stable generation of \unit{400}{mW} at \unit{470}{nm} starting from \unit{560}{mW} at \unit{940}{nm}. In the second stage, with \unit{140}{mW} input to the cavity, an output power of \unit{28}{mW} at \unit{235}{nm} is obtained. For quantum control of beryllium ions, three laser wavelengths at \unit{313}{nm} are produced by sum-frequency generation and subsequent SHG, starting from four infrared fiber lasers. Up to \unit{7.2}{W} at \unit{626}{nm} have been generated from \unit{8.5}{W} at \unit{1051}{nm} and \unit{8.3}{W} at \unit{1551}{nm}. The red light is then frequency-doubled in bowtie cavities using BBO crystals. With an incident power of \unit{3.8}{W}, we obtain \unit{1.9}{W} of UV light at \unit{313}{nm}. We have tested these systems by successfully loading \Beplus ions in a Paul trap.

The quantum information experiments we envisage will probably benefit largely from these new systems. Photoionization of $^9$Be atoms with the \unit{235}{nm} light should be more efficient than ionization with an electron gun. Although a CW laser of such short wavelength can easily lead to electrically charging the trap electrodes, we find that we can load \Beplus ions reliably in the trap after careful shaping and positioning of the beam. On the other hand, the power delivered by the \unit{313}{nm} systems will help in the long-term goal of fault-tolerant quantum computation using stimulated Raman transitions, since the difficulties derived from the trade-off between spontaneous scattering and quantum-gate speed can be overcome with high-intensity laser beams.

\begin{acknowledgements}
We thank Florian Leupold (ETH Z\"urich) for careful reading and comments on the manuscript, Andrew Wilson and Dietrich Leibfried (NIST, Boulder) for useful information on cavity design and helpful discussions, and Christian Rahlff (Covesion Ltd.) for information on PPLN crystals. This work was supported by the Swiss NSF under grant no. 200021 134776, the NCCR QSIT and ETH-Z\"urich.
\end{acknowledgements}

\appendix

\section{Second-harmonic-generation efficiency}

Here we give the expressions used for the theoretical calculation of the second-harmonic-generation efficiencies plotted in figures \ref{fig:conveffvsPf}, \ref{fig:conveffvsPs} and \ref{fig:SHG_Raman_Power}. The values for the relevant parameters are given in table \ref{tab:gammatheo}.

We start from the original equations from Boyd and Kleinman \cite{68Boyd} and rearrange them to formulate the conversion efficiency (in SI units \cite{BkRisk}) as
\be
\nonumber \Gamma_\text{eff}=\frac{P_{2\omega}}{P_\omega^2}= & \frac{16\pi^2d_\text{eff}^2}{\lambda^3\epsilon_0cn_\omega n_{2\omega}}l_\text{c}e^{-\rbra{\beta_\omega+\frac{\beta_{2\omega}}{2}}l_\text{c}}\times\\
 & h(b,l_\text{c},\beta_\omega,\beta_{2\omega},\sigma,B).
\ee
Here, $P_{2\omega}$ is the SHG power and $P_{\omega}$ the power in the fundamental wavelength at the crystal (the circulating power if a cavity is used); $d_\text{eff}$ is an effective non-linear coefficient in units of m$\cdot$V$^{-1}$ (related but not identical to parameter $d$ in \cite{68Boyd}); $\epsilon_0$ and $c$ are the vacuum permittivity and speed of light, respectively; $\lambda$ is the pump wavelength, $l_\text{c}$ the crystal length, $\beta_\omega$ ($\beta_{2\omega}$) the absorption coefficient of the fundamental (second harmonic) light in the crystal, and $n_\omega$ ($n_{2\omega}$) the refractive index of the crystal sampled by the light in the fundamental (second harmonic) wavelength.

$h$ is a dimensionless function which depends on $l_\text{c}$, $\beta_\omega$, $\beta_{2\omega}$, the confocal parameter $b=2\pi n_\omega w^2/\lambda$ (where $w$ is the waist radius of the pump beam, assumed to be at the center of the crystal), the wavevector mismatch $\sigma$ (scaled by $b$) and the walk-off parameter $B=\rho(\theta,\lambda)\sqrt{\pi l_\text{c}n_\omega/(2\lambda)}$ where $\theta$ is the phase-matching angle and
\be
\rho(\theta,\lambda)=\arctan\rbra{\frac{1-\rbra{\frac{n_{\text{o,}2\omega}}{n_{\text{e,}2\omega}}}^2}{\cot\theta+\rbra{\frac{n_{\text{o,}2\omega}}{n_{\text{e,}2\omega}}}^2\tan\theta}}.
\ee
$n_{\text{o,}2\omega}$ ($n_{\text{e,}2\omega}$) is the refractive index for the second-harmonic light along the ordinary (extraordinary) crystal axis.

The explicit expression for $h$ is,
\be
\nonumber h & = &\frac{1}{4\xi}\int_{-\xi}^{\xi}\D\tau'\int_{-\xi}^{\xi}\D\tau\frac{1}{(1+\I\tau)(1-\I\tau')}\times\\
\nonumber & & \text{exp} \bigg\{-\frac{b}{2}\rbra{\beta_\omega-\frac{\beta_{2\omega}}{2}}(\tau+\tau')+\\
 & & \I\sigma(\tau-\tau')-\frac{B^2}{\xi}(\tau-\tau')^2\bigg\},
\ee
with $\xi=l_\text{c}/b$.

In a cavity the relation between the circulating power $P_\omega$ and the fundamental power $P_{\omega\text{,0}}$ pumped into it is given by \cite{05LeTargat}
\be
\frac{P_\omega}{P_{\omega\text{,0}}}=\frac{T_1}{\rbra{1-\sqrt{(1-T_1)(1-\epsilon)(1-(\Gamma_\text{eff}+\Gamma_\text{abs})P_\omega)}}^2},
\ee
where $T_1$ is the transmission of the in-coupling mirror, $\epsilon$ the round-trip loss in the cavity (excluding frequency conversion) and $\Gamma_\text{abs}$ the absorption efficiency of the frequency-doubled light. If this absorption takes place only in the crystal and we consider the limit that frequency conversion occurs exclusively in its center, we can make the approximation
\be
\Gamma_\text{abs}\approx\Gamma_\text{eff}\rbra{\Exp{\beta_{2\omega}l_\text{c}/2}-1}.
\ee

The power ratio given in figures \ref{fig:conveffvsPf}, \ref{fig:conveffvsPs} and \ref{fig:SHG_Raman_Power} is the ratio between the net power of the second-harmonic light at the output of the doubling cavity $P_{2\omega\text{,out}}$ and the fundamental power at the cavity input:
\be\label{eq:eta}
\eta_\lambda\equiv\frac{P_{2\omega\text{,out}}}{P_{\omega\text{,0}}}=\frac{P_{2\omega}(1-R_{2\omega})}{P_{\omega\text{,0}}}=\frac{P_{\omega}^2\Gamma_\text{eff}(1-R_{2\omega})}{P_{\omega\text{,0}}},
\ee
where $R_{2\omega}$ is the reflection of the second-harmonic light at the output surface of the crystal. For the case of QPM in PPKTP and PPLN the second-harmonic light propagates perpendicularly to the surface, so there is no reflection. However, in the 470$\rightarrow$\unit{235}{nm} and 626$\rightarrow$\unit{313}{nm} cavities, the incidence is not normal to the (Brewster-cut) surface and the polarization of the second-harmonic light is along the vertical direction, so $R_{2\omega}>0$ and not all the power generated in the doubled frequency is available at the cavity output. Equation \ref{eq:eta} can be solved with the expressions above given the cavity-related parameters $T_1$, $\epsilon$ and $w$; the crystal parameters $l_\text{c}$, $d_\text{eff}$, $n_{\omega,2\omega}$, $\beta_{\omega,2\omega}$ and $R_{2\omega}$; the phase-matching parameters $\sigma$ and $\theta$; and the properties of the light matched into the cavity, $\lambda$ and $P_{\omega,0}$.

\begin{table*}\scriptsize
\caption{Values used for calculating the expected conversion efficiency from the frequency-doubling cavities. $d_\text{eff}$ is the effective non-linear coefficient of the crystals (for BBO, see \cite{OlCastechBBO}). Perfect phase matching is assumed in all cases, no absorption inside the BBO crystals and no walk-off inside the PPKTP.}
\label{tab:gammatheo}       
\begin{center}
\begin{tabular}{c c c c}
\hline
& 940$\rightarrow$470 nm & 470$\rightarrow$235 nm & 626$\rightarrow$313 nm \\
\hline\hline
 $l_\text{c}$ & \unit{30}{mm} & \unit{10}{mm} & \unit{10}{mm} \\
 $d_\text{eff}$ & \unit{9.5}{pm/V} & \unit{1.45}{pm/V} & \unit{2.08}{pm/V} \\
 $n_\omega/n_{2\omega}$ & 1.8361/1.9149 & 1.6810/1.6408 & 1.6676/1.5888 \\
 $w$ & \unit{50}{\micro m} & \unit{17.9}{\micro m} & \unit{20.7}{\micro m} \\
 $\beta_\omega/\beta_{2\omega}$ & \unit{0.3/14}{m^{-1}} & 0 & 0 \\
 $R_{2\omega}$ & 0 & \unit{22}{\%} & \unit{20}{\%} \\
 $\sigma$ & 0 & 0 & 0 \\
 $B$ & 0 & 18.9 & 16.4 \\
 $\epsilon$ & \unit{2}{\%} & \unit{0.7}{\%} & \unit{0.9}{\%} \\
 $T_1$ & \unit{14.5}{\%} & \unit{0.9}{\%} & \unit{1.6}{\%} \\
\hline
\end{tabular}
\end{center}
\end{table*}

\bibliographystyle{spphys}       
\bibliography{myrefs}

\end{document}